\documentclass[twocolumn,aps,prb]{revtex4}

\usepackage{amsmath}
\usepackage{amssymb}
\usepackage{graphicx}
\usepackage{epsfig}
\usepackage{bm}
\usepackage{units}
\usepackage{subfigure}
\usepackage{hyperref}
\usepackage{breakurl}

\newcommand{\be}{\begin{equation}}
\newcommand{\ee}{\end{equation}}
\newcommand{\bea}{\begin{eqnarray}}
\newcommand{\eea}{\end{eqnarray}}
\newcommand{\ben}{\begin{enumerate}}
\newcommand{\een}{\end{enumerate}}
\newcommand{\bit}{\begin{itemize}}
\newcommand{\eit}{\end{itemize}}

\begin{document}

\title{On the Title of Moriarty's {\it Dynamics of an Asteroid}}
 
\author{Alejandro Jenkins}\email{ajenkins@fisica.ucr.ac.cr}

\affiliation{Escuela de F\'isica, Universidad de Costa Rica, 11501-2060, San Jos\'e, Costa Rica}
\affiliation{High Energy Physics, Florida State University, Tallahassee, FL 32306-4350, USA}

\date{Feb.\ 2013, last revised Aug.\ 2014}

\begin{abstract}

We propose an explanation of the title of Prof.\ James Moriarty's treatise {\it Dynamics of an Asteroid}, a scientific work mentioned by Sherlock Holmes in {\it The Valley of Fear} and prominently featured in Guy Ritchie's 2011 film {\it Sherlock Holmes: A Game of Shadows}.  Our views on the subject differ from those expressed in Isaac Asimov's ``The Ultimate Crime.''

\end{abstract}

\maketitle



Isaac Asimov\cite{Asimov} and others\cite{Schaefer} have remarked on the peculiarity of the title of Prof.\ James Moriarty's book {\it The Dynamics of an Asteroid}, a scientific treatise that, in the words of Sherlock Holmes, ``ascends to such rarefied heights of pure mathematics that it is said that there was no man in the scientific press capable of criticizing it.''\cite{Fear}  In particular, why is {\it asteroid} used in the singular?  Asimov had his own ideas about this, but there might be a more plausible solution to this puzzle.

In Victorian Britain there were several textbooks called {\it Dynamics of a Particle}.  For example, Peter Guthrie Tait and William J.~Steele collaborated on a textbook of that name, intended for Cambridge undergraduates.  That work appeared in 1856 and went through seven editions, the last from 1900.\cite{Tait}  There is also a {\it Dynamics of a Particle} by Edward Routh, published in 1898.\cite{Routh}  A search of the Harvard library catalogue returns several subsequent publications with similar titles, by R.~J.~A.~Barnard,\cite{Barnard} S.~L.~Loney,\cite{Loney} and W.~D.~MacMillan.\cite{MacMillan}  Why was {\it particle} used there as a singular noun?

In the scientific parlance of the time, ``particle'' meant something rather different from the sense modernly attached to the word by quantum physics: it referred to a solid body of fixed mass, whose physical state at a given time may be entirely characterized by one position and one velocity.  In particular, the actual size and shape of the body are irrelevant, so that no rotation or any other internal motion or property need be taken into consideration.  A synonym for particle in this sense is ``material point.''\cite{Landau}

By considering the motion of a single particle, the student avoids the complications introduced by the changing interactions between particles as they move relative to one another.  Thus, in Moriarty's day, ``dynamics of a particle'' was a standard first course in mathematical physics, covering essentially the same material as a modern introductory course in Newtonian mechanics, like the one that most university students in the natural sciences are required to complete today.  In the Victorian physics curriculum, this would have been followed by more advanced studies on the dynamics of systems of particles, of rigid bodies, of elastic solids, and of fluids.

Gauss and other 19th-century mathematical scientists who worked on the subject treated an asteroid as a particle, subject only to the gravitational attraction of the Sun and to small perturbations from the influence of nearby planets.\cite{Gauss} In this context, the title {\it Dynamics of an Asteroid} suggests that Moriarty's approach was general and theoretical, closer to pure mathematics than to observational astronomy.  This is the opposite of Asimov's interpretation, who concluded that Moriarty must have had a specific asteroid in mind.  It is, on the other hand, a view strongly supported by the fact that Moriarty's other known publication was his youthful paper on the binomial theorem, a strictly mathematical subject.\cite{Anderson}

Mathematics in 19th-century British universities was still under Newton's influence, so that the dynamics of a particle (such as an asteroid) would have been a subject of interest to mathematicians just as much as to physicists.  Lewis Carroll, creator of {\it Alice in Wonderland} and ---as Charles L. Dodgson--- mathematical lecturer at Oxford, published in 1865 a satirical pamphlet called ``The Dynamics of a Parti-cle'' ({\it sic}), which dealt facetiously with certain issues of the Oxford politics of the day, especially William Gladstone's defeat in his bid for reelection as Member of Parliament for the university.\cite{Carroll}  Carroll's title is evidently a play between the name of an introductory course in mathematical physics, with which Dodgson and his colleagues would have been very familiar, and the political sense of the world ``party.''

Moriarty might have named his treatise on celestial mechanics by analogy to an introductory physics text in order to encourage students to read it.  In this he must have failed, given what we know of the work's mathematical abstruseness.  It is not uncommon for great theoretical scientists to underestimate the mathematical difficulties that their work will pose for common readers.  For instance, Sir Roger Penrose's {\it Road to Reality}, published in 2004, is intended for a lay audience but includes discussions of hypercomplex numbers, symplectic manifolds, Riemann surfaces, and gauge connections, among many other topics in higher mathematics.\cite{Penrose}

\begin{figure} [t]
\begin{center}
	\includegraphics[width=0.42 \textwidth]{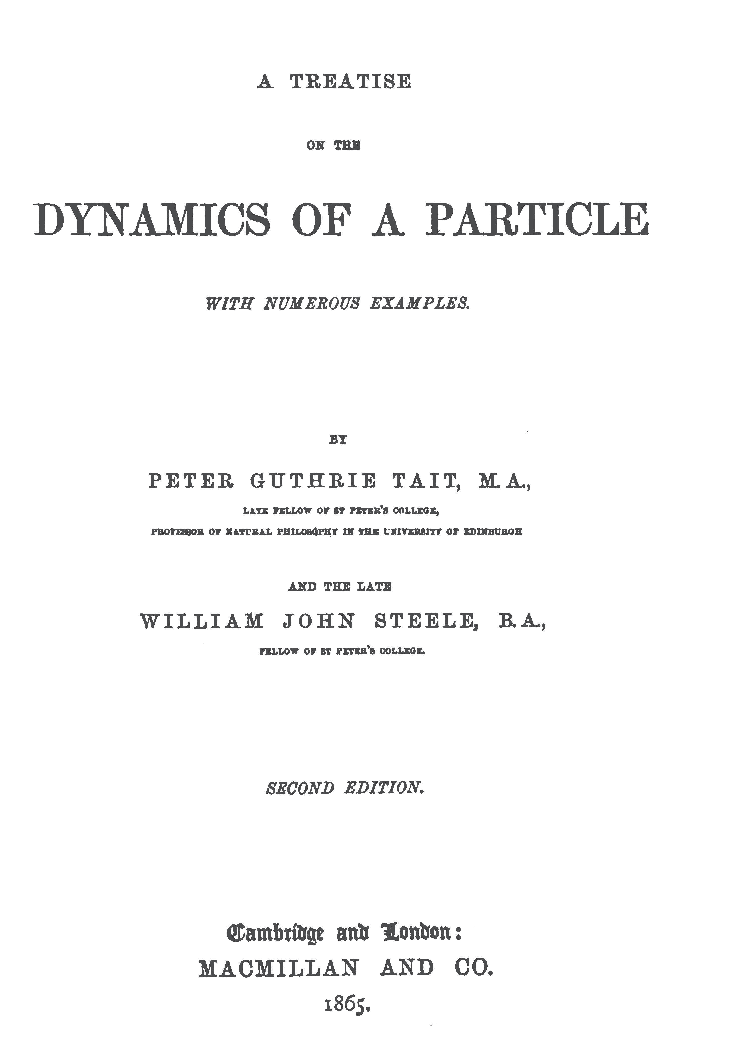}
\end{center}
\caption{\small Title page of the second edition of {\it Dynamics of a Particle} by Tait and Steele.}
\end{figure}

Asimov argued that the study of the motion of a generic asteroid, treated as a particle, would have been a well-worn subject by 1875 (around which time he estimated that Moriarty's work was written), and therefore would have afforded little scope for the author's genius.\cite{Asimov}  But we must not forget that Cauchy's work on complex-valued functions, a deathless {\it tour de force} of pure mathematics, grew out of his study of Kepler's equation for the elliptical orbit of a planet going around the Sun.  That very same problem had already inspired Newton to invent topology, an entirely new branch of mathematics, but that work was so far ahead of its time that it lay forgotten for 300 years.\cite{Arnold}

In 1890, Poincar\'e tackled the difficult and longstanding problem of characterizing mathematically the motion of three celestial bodies as they pull gravitationally on each other.\cite{Gutzwiller}  That work inaugurated what would much later be dubbed ``chaos theory,'' a subject that began to attract the attention of both scientists and the general public after digital computers made it possible to plot complicated trajectories that were otherwise very difficult to calculate and visualize.\cite{Holmes}

A treatment like Poincar\'e's of the three-body interaction would probably have been beyond the scope of a book on the ``dynamics of an asteroid.''  On the other hand, it was established mathematically in the 1980s that an asteroid's orbit may, under certain circumstances, become chaotic due to the recurring gravitational tug of a nearby planet.  An asteroid in such a chaotic orbit is likely to collide against a planet eventually.  This explains the dearth of objects in certain narrow regions of the asteroid belt, known as ``Kirkwood gaps,'' whose existence had already been noticed in Moriarty's day.\cite{Kirkwood}

This raises the possibility (suggested to me by Marshall Eubanks) that Prof.\ Moriarty, the ``Napoleon of crime,''\cite{Napoleon} might be the unacknowledged founder of the mathematical theory of chaos.  Surely the subject would have appealed to his diseased genius.  Could such work have informed Moriarty's development of the most advanced criminal network of his day?

In any case, we may surmise that Moriarty's study of the motion of an asteroid drove him to develop original mathematical concepts, much as Newton, Gauss, Cauchy, and Poincar\'e did in the course of their own researches in celestial mechanics.  Moriarty's work was evidently not understood at the time, and unfortunately it was later lost, probably because it was suppressed after the author's criminal career became widely known.

\begin{acknowledgements}

Thanks are due to several readers who commented on this manuscript after it first appeared on the physics preprint archive.  Marshall Eubanks offered a compelling argument that Moriarty's book could have anticipated some aspects of Poincar\'e's work on chaos.  Jennifer Ouellette discussed this manuscript on her blog (``Cocktail Party Physics,'' hosted by {\it Scientific American}), providing a venue for further reflection on the connections between Sherlock Holmes, mathematics, and astronomy.\cite{Ouellette}  The author also thanks Vanessa L\'opez, his analytical mechanics students at the University of Costa Rica, and the listeners of his TEDxJoven@PuraVida 2013 talk (``{\it Crimen y astronom\'ia: El genio malvado del Prof.\ Moriarty'}') for their feedback.

\end{acknowledgements}


\bibliographystyle{aipprocl}   

\end{document}